# Hybrid Context Retrieval Augmented Generation Pipeline: LLM-Augmented Knowledge Graphs and Vector Database for Accreditation Reporting Assistance

**Candace Edwards**

Information and Computer Science, University of Hawaii at Manoa
ICS 699: MS Plan B Capstone Report
April 21, 2024
cedward2@hawaii.edu

**Abstract**

In higher education, accreditation is a quality assurance process, where an institution demonstrates a commitment to delivering high quality programs and services to their students. For business schools nationally and internationally the Association to Advance Collegiate Schools of Business (AACSB) accreditation is the gold standard. For a business school to receive and subsequently maintain accreditation, the school must undertake a rigorous, time consuming reporting and peer review process, to demonstrate alignment with the AACSB Standards. For this project we create a hybrid context retrieval augmented generation pipeline that can assist in the documentation alignment and reporting process necessary for accreditation. We implement both a vector database and knowledge graph, as knowledge stores containing both institutional data and AACSB Standard data. The output of the pipeline can be used by institution stakeholders to build their accreditation report, dually grounded by the context from the knowledge stores. To develop our knowledge graphs we utilized both a manual construction process as well as an 'LLM Augmented Knowledge Graph' approach. We evaluated the pipeline using the RAGAs framework and observed optimal performance on answer relevancy and answer correctness metrics.

**Project Overview[1]:**

This project implements an hybrid-context retrieval augmented generation architecture, integrating both data stored as vector embeddings and data structured as a knowledge graph. This project implements several query optimization techniques and multi-source retrieval to provide context to ground the generated output response. This project uses Large Language Models for several tasks including knowledge graph construction.

The project is built using Python and Cypher, with Neo4j as the database management system (DBMS). The project also combines frameworks like LangChain, with custom pipeline development. Along with the RAGAs framework for pipeline evaluation.

**Introduction:**

Accreditation maintenance for degree granting higher education institutions is a multi-year task involving dozens of internal and external stakeholders. For business schools nationally and internationally the Association to Advance Collegiate Schools of Business (AACSB) accreditation is the gold standard for accreditation. The accreditation signifies a commitment to: engaging curriculum, student success, valuable research contribution, cutting edge facilities and more. Over 900 business schools across the world are AACSB accredited.

As the current protocol stands, schools are required to submit a 'Continuous Improvement Report' and prepare for an AACSB on-site visit every five years. The development of this report and preparation for the peer review visit can take about a year of preparation. The school's are evaluated based on their alignment with the AACSB Standards.

---

[1] GIT: https://github.com/CS-Edwards/advRAG
BRANCH: Accreditation

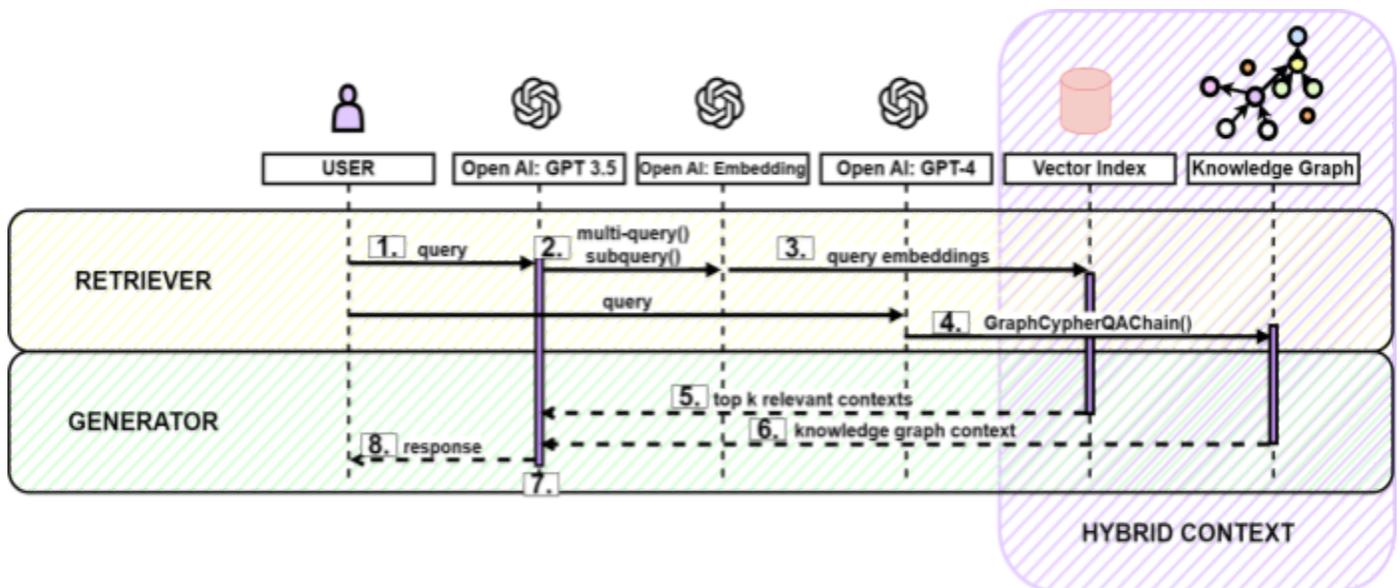

Figure 1. System Sequence Diagram. 1 .User Query: User initiates natural language query. 2. Query Optimization: Query is routed and expanded to multiple queries. 3. Query Embedding: Queries are passed to the text embedding model. 4. Query Conversion: Original natural language query is passed into model, and converted to a Cypher query using LangChain's GraphCypheryQAChain(). 5. Vector Index Context Retrieval: Vector index returns the top k (k=2), embeddings based on cosine similarly to the query embeddings. 6. Knowledge Graph Context Retrieval: Knowledge graph query returns information based on the Cypher query. 7. Generator: Hybrid context from vector index and knowledge graph are passed in the model along with the original query to generate response to the query. 8.Response Returned: Generated response is returned to the user.

The goal of this project is to implement an hybrid-context retrieval augmented generation pipeline, for the purposes of easing the report development process. Utilizing this pipeline, institutions will be able to integrate their own documents into the database which is prepopulated with AACSB Standard data and users will be able to query their own institutional data, alongside AACSB data, to evaluate how their documents align with the standards.

For example, a school may host an student entrepreneurship event which is captured in documentation by way of an article in the campus or local news paper. And additionally hold a curriculum assessment committee meeting where student performance on a learning objective is reviewed, documented by the minutes of the meeting. In this example the news article, and meeting minutes are added to the pipeline and aligned (linked) to their relevant standards. The entrepreneurship event news article demonstrates alignment with Standard 6: Learner Progression, and the committee minutes align with Standard 5: Assurance of Learning.

With this data and other institutional documents stored in the pipelines knowledge store, a user can input a query ie:

"How did we demonstrate alignment with AACSB Standard 6."

And the pipeline would return a natural language response ie:

"To demonstrate alignment with AACSB Standard 6: Learner Progression, we showcased our commitment to student entrepreneurship, through our annual 'Student Innovators' event…"

Note the pipeline would collect additional relevant examples in the knowledge store and include them in the response.

With the assistance of the pipeline, the time consuming and manual process of searching through documentation to aggregate relevant demonstration of standards, and subsequently drafting a written report based on documents is expedited. The pipeline output provides a more efficient starting point for the reporting process.

Given that the queries use natural language, the pipeline makes institutional accreditation accessible to a wide range of stakeholders including, administrators, faculty, students, and committees.

**System Overview:**

In this section we provide the Hybrid Context RAG discussing the sequence and flow of a user query through the pipeline. Each step is explained in detail throughout this report. This sequence is illustrated in Figure 1.

**System Flow Steps:**

1. **User Query:** User initiates natural language query.

   $q^0$: "Which learning objectives did out undergraduate program evaluate"

2. **Query Optimization:** Query is routed and expanded to multiple queries.

   $q^1$: "What are the learning objectives that our undergraduate program evaluated?"
   $q^2$: "What were the learning objectives assessed in our undergraduate program?"
   $q^3$: "Which learning objectives were evaluated in our undergraduate program?"

3. **Query Embedding:** Queries $\{q^1, q^2, q^3\}$ are passed to the text embedding model.

4. **Query Conversion:** Original natural language query ($q^0$) is passed into model, and converted to a Cypher query using LangChain's GraphCypheryQAChain().

5. **Vector Index Context Retrieval:** Vector index returns the top k (k=2), embeddings based on cosine similarly to the query embeddings.

6. **Knowledge Graph Context Retrieval**: Knowledge graph query returns information based on the Cypher query. In Figure 2, we illustrate the node view of the query result. The natural language context result is returned.

7. **Generator:** Hybrid context from vector index and knowledge graph are passed in the model along with the original query to generate response to the query.

8. **Response Returned:** Generated response is returned to the user.

**Background:**

In this section we provide our grounding research on the key interlocking elements of our project: Retrieval Augmented Generation (RAG), Knowledge Graphs (KGs) and Large Language Models (LLMs).

**Retrieval Augmented Generation:**

Large Language Models have achieved mainstream popularity, beyond the tech industry and research community, because of their ability to generate human-like completions to user inputs.

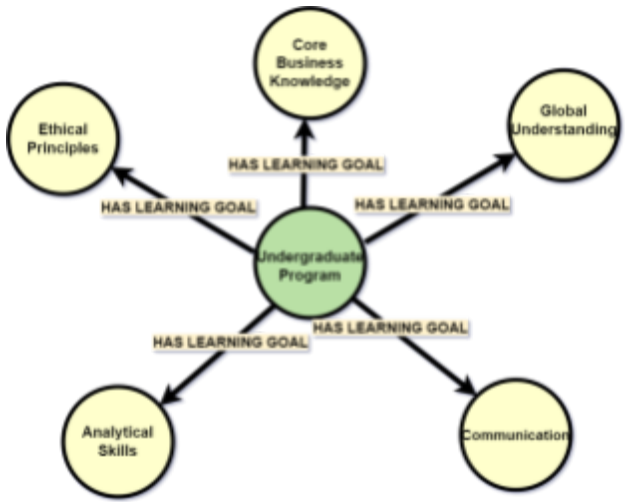

Figure 2. Node level view of knowledge graph query result. Original query: "Which learning objectives did our undergraduate program evaluate". 'Undergraduate Program' node, linked to 'Learning Goal' nodes (Global Understanding, Communication, Analytical Skills, Ethical Principles, Core Business Knowledge), using the relationship label "HAS_LEARNING_GOAL".

A language model (LM), is a model that 'looks to estimate the probability distribution over text' [2]. As language models have grown in model parameter size (from millions to hundreds of billions) with larger training data (corpra), we arrive at Large Language Models (LLMs). Based on their pre-training and fine-tuning, LLMs excel at many natural language tasks including text-generation [3], summarization[3] and question-answering[4]. However, LLMs have several limitations including: hallucinations [5] – where the model generates a response that is factually inaccurate, knowledge limitations[6] confined by the training copra, lack of transparency and untraceable reasoning processes [7].

According to a 2023 study by the Pew Research Center [1], approximately 58% of adults in the U.S. have at least heard of ChatGPT, OpenAI's proprietary chat assistant and most popular tool – currently based on their gpt-3.5-turbo model.

Retrieval Augmented Generation (RAG) is a system architecture approach that allows for Large Language Models (LLMs) to leverage external contextual information to generate responses to input prompts [6]. The RAG architecture overcomes several limitations of stand-alone LLMs. Y.Gao et al.[7] summarize current RAG approaches into three paradigms: Naive RAG, Advanced RAG and Modular RAG.

*Naive RAG:*

The foundational and most basic implementation of the RAG method. Naive RAG begins with splitting raw text data that may contain interaction-relevant into smaller 'chunks', encoded into a vector representation and placed into a knowledge store (eg. vector database). The *retrieval* step takes a user query and encodes the query into a vector representation and performs a semantic search against the indexed vectors to return chunks with the highest (top-k) similarity scores. The *generation* step takes the retrieved chunks as context and pairs them with the user's query to pass into an LLM as a prompt to generate a relevant response. The challenges of the Naive RAG approach include retrieval issues where returned vectors are missing crucial information or are non-relevant to the query, and generation issues, including hallucinations where generated responses are not based on the provided context. "Facing complex issues, a single retrieval based on the original query may not suffice to acquire adequate context information" [7].

*Advanced RAG:*

The Advanced RAG method extends the Naive RAG approach to address retrieval limitations. Advanced RAG adds two process steps to improve the retrieval process. Pre-retrieval focuses on optimizing the indexing structure and the original query. Both index optimization and query optimization will be discussed later. The second added process step consists of re-ranking the retrieved contexts based on relevance to the original query. This added step aims to "concentrate on selecting essential information, emphasizing critical sections and shortening the context to be processed"[7].

*Modular RAG:*

Modular RAG extends beyond both Naive and Advanced RAG by providing additional functionality in the form of modules and allows for a flexibility of architecture pattern which can improve the system. Such modules include "Search", which enables search across multiple data sources like databases and knowledge graphs. "Predict", a module with a 'generate-then-read'[8] approach where an LLM is prompted to first generate contextual documents *instead* of retrieving documents from an external knowledge source. One innovative Modular RAG pattern is the 'Demonstrate-Search-Predict' framework, where the 'Demonstrate' phase provide a train example as a form of 'weak supervision', 'Search' gathers information from a knowledge base and 'Predict' generates grounded output [9].

The design of our pipeline is the most aligned with the Advanced RAG approach. We implement query and index optimization along with context retrieval from multiple sources. Our system is more complex than the Naive RAG model, and does not incorporate the modularity of the Modular RAG pattern.

Index and Query Optimizations:
As noted earlier, Advanced RAG, focuses on efficient retrieval of relevant documents based on the input query. Here explore both indexing optimization and query optimization considerations.

Index Optimization
The indexing phase ingests the project specific documents that will be used to provide the external knowledge to the system and stores the documents in the knowledge store (eg. vector database, knowledge graph, triple store,etc). Within index optimization there are three areas to be considered to enhance the index process: 1. chunking strategy, 2. metadata attachments and the 3. structural index.

*1.Chunking Strategy*

Chunking is the task of splitting whole documents into smaller, more manageable segments. There are several chunking methods to be considered [10]. *Fixed size* chunking, splits the document based on a fixed number of tokens. While this is the simplest approach, it is not ideal since the split does not maintain the context of the within the document (eg. potential to split a sentence or even a token). *Chunk Overlap* provides some continuity of context, where chunks contain a predefined portion of text from their preceding chunk. *Sentence splitting* and *recursive chunking* are examples of 'context-aware' chunking strategies, which attempt to keep relevant lines together. There is also *semantic* chunking [11,12], which focuses on grouping together sentences that are semantically related.

*2. Metadata Attachments*
Metadata attachments refers to enriching chunks with metadata information (eg. page number, file name, timestamp, etc). Chunks can then be filtered based on their metadata, further refining the retrieval and improving relevance to the query [7].

*3. Structural Index*
Adding a hierarchical structure to documents in the index aids in 'swift traversal' of the data, where files are arranged in parent-child relationships. Implementing a Knowledge Graph index captures logical relationships between the document, content and structure [13]. More on knowledge graphs in subsequent sections.

*Query Optimization*
The context collected by the retriever and response generated by the LLM is directly dependent on the input query. As such, a poor query (eg. imprecise, unclear) will result in a sub-par generation. Query optimization seeks to enhance the effectiveness and quality of the input query. Within query optimization there are three areas to be considered: 1.query expansion, 2. query transformation and 3. query routing.

*1. Query Expansion*
Query expansion expands a single query into multiple queries to enhance query quality. The approaches to query expansion are: multi-query, sub-query and Chain-of-Verification(CoVe). Muli-query utilizes prompt engineering to generate multiple questions based on the initial input query [14]. Suq-query utilizes prompts to break a complex query into simple sub-questions. The CoVe method attempts to reduce hallucinations by validating expanded queries [15].

*2. Query Transformation:*
The query transformation approach completely rewrites the user query and retrieves chunks based on the modified query.

*3. Query Routing:*
Query routing evaluates the query to then route to a 'distinct RAG pipeline', where the query can be routed by metadata based on extracted keywords, or routed semantically. These approaches are not mutually exclusive; hybrid routing can also be employed to enhance the query.

The data structure of the knowledge store is also an important factor in overall retrieval performance. We have briefly mentioned vector databases as a knowledge store for RAG architectures. In the next section we discuss at Knowledge Graphs at length, a structured knowledge store also used in RAG implementations.

**Knowledge Graphs:**

A knowledge graph (KG), also known as a semantic network, is a structure that depicts how concepts are related to one another. A knowledge graph contains three main elements: nodes, edges and edge labels [16], where the node represents an entity, the edge represents the relationship between two nodes and the label defines the relationship.

By inherently maintaining the relationship structures between entities, knowledge graphs facilitate the representation of complex interconnections within the domain of data contained in the graph. This structured representation enables effective data integration, information retrieval, and knowledge inference; all while providing human-readable structure.

Knowledge graphs have applications across multiple domains including education [17] – the domain for this project.

The term 'knowledge graph' was popularized by the success of large scale projects such as the Google Knowledge Graph in 2012 [18]. M.Kejriwal (2022) reviews the study and development of knowledge graphs across multiple communities including: Natural Language Processing (NLP), Semantic Web (SW) , Machine Learning (ML), Knowledge Discovery Databases (KDD) and data mining; identifying a unified system level overview of the components of a knowledge graph across domains [19]. M. Kejriwal [19] describes the knowledge graph development sequence is as follows:

1. Domain Modeling and Data Acquisition: Collecting and compiling data (structured, semi-structured and/or unstructured) and developing the ontology.
2. Knowledge Graph Construction and Identification: The process of building the knowledge graph from data
3. Knowledge Graph Access: Placing the identified graph in a "querying and storage architecture" for use.

*Challenges: Knowledge Graph Construction*
While a major benefit of knowledge graphs is their inherent readability and interpretability [20], the knowledge graphs themselves are challenging to construct [21,22]. M.Hofer et al. [21], outline the construction tasks as:

● **Data Acquisition & Preprocessing:** Knowledge Completion: Extending a given KG, e.g., by learning missing type information, predicting new relations, and enhancing domain-specific data (polishing).
● **Metadata Management:** Acquisition and management of different kinds of metadata

- **Ontology Management:** Creation and incremental evolution of KG ontology
- **Knowledge Extraction (KE)**: Derivation of structured information and knowledge from unstructured or semistructured data. using techniques for named entity recognition, entity linking and relation extraction.
- **Entity Resolution (ER) and Fusion:** Identification of matching entities and their fusion within the KG.
- **Quality Assurance (QA):** Possible quality aspects, their identification, and repair strategies of data quality problems in the KG.
- **Knowledge Completion:** Extending a given KG, e.g., by learning missing type information, predicting new relations, and enhancing domain-specific data (polishing).

As noted above there are several steps involved in the construction of a knowledge graph, where decisions on how to approach each step are dependent on the data and use case; making a generalized solution/ approach intractable. Another major challenge [21] is the difficulty in integrating new information into the knowledge graph once it has been constructed. As stated for the previous section, we will detail the knowledge construction pipeline development and challenge resolution approaches for our specific use case in the methodology section. In the next section we'll explore how Large Language Models (LLMs) can be used to mitigate some of the knowledge graph construction challenges along with other synergies for LLMs and KGs.

*Large Language Models and Knowledge Graphs*

In the earlier discussion on Retrieval Augmented Generation (RAG) where knowledge graphs serve as a knowledge source, we note that Large Language Models (LLMs) are able to effectively combine user query with retrieved context knowledge into a human-like response. This symbiotic relationship between knowledge graphs and LLMs also extends beyond the RAG use case [23, 24].

S.Pan et al. [24] layout a comprehensive 'roadmap' on the mutually beneficial relationship between LLMs and Knowledge Graphs, devising a framework of three broad categories of LLM/KG collaboration structures:

- **KG-Enhanced LLMs:** Integrating LLMs with KGs to enhance the performance and interpretability of LLMs in various downstream tasks. This category consists of three groups: KG-Enhanced LLM pre-training, KG-Enhanced LLM Inference and KG-enhanced LLM interpretability.
- **LLM-Augmented KGs:** Applying LLMs to augment KG related tasks. This category consists of five groups: LLM-augmented embedding, LLM-augmented KG completion, LLM-augmented KG construction, LLM-augmented KG-to-text Generation and LLM-augmented KG question-answering
- **Synergized LLMs+ KGs:** Mutual enhancement of LLMs and KGs from "the perspectives of knowledge representation and reasoning"

The RAG structure of this project falls within the category of "KG-Enhanced LLMS", specifically within the subsection of "KG Enhanced LLM Inference: Retrieval-Augmented Knowledge Fusion" where "the key idea is to retrieve relevant knowledge from a large corpus and then fuse the retrieved knowledge into LLMs"[24]. We have established this already.

We now turn to the second category, "LLM-augmented KG's", to address some of the challenges raised earlier in our discussion of knowledge graph construction. M.Kejriwal highlights Information Extraction (IE), Named Entity Recognition (NER) extraction, and Relationship Extraction (RE) as the NLP tasks associated with knowledge graph research and development [19]. Large Language Models can be used to facilitate these NLP tasks. S.Pan et al .[24], directly address the Knowledge Extraction (KE), Entity Resolution (ER) and Fusion, tasks of knowledge graph construction identified earlier. In the proposed framework "LLM-Augmented Graph Construction" consists of five sub categories: entity discovery, coreference resolution, relation extraction, end-to-end KG construction and distilling KGs from LLMs. Below we outline the first four since they are relevant to this project.

Entity Discovery is the "process of identifying and extracting entities from unstructured data sources"[24]. This process consists of Name Entity Recognition (NER), the task of recognizing and categorizing named entities (ie. people, organizations, locations, etc) in a text document [25], Entity Typing (ET), which provides context to the recognized entity (eg. AACSB is the entity, type is organization), and Entity Linking (EL) also known as entity disambiguation or Coreference Resolution [24] , which connects text entities to their corresponding nodes in the knowledge graph Note, M.Hofer et al. refer to this task as Entity Resolution (ER) and Fusion.

Coreference Resolution (CR) is "to find all expressions that refer to the same entity or event in a text" [24], and consists of two subtasks: Within-document CR and Cross Document CR.

Relation Extraction (RE) "involves identifying semantic relationship between entities in natural language text"[same doc], consisting of two methods: Sentence-Level RE and Document-Level RE, where the former identifies entity relations within an sentence and the latter identifies entity relationships between sentences within the same document.

We discuss our approach to 'LLM Augmented KG Construction' later in this report.

**Models and Data Overview:**

In this section we review the data and models used in this project. Along with the tasks for which the models are used.

Data Summary:
The AACSB Accreditation Standard data is sourced directly from the organization website[2]. In this project we used the 2020 version of the accreditation standards – the version of standards currently in use. As discussed in the project overview, the standards are the foundation for each use case. As a user facing product, the accreditation standard data would be preloaded on the database for each client. The accreditation standards were originally downloaded from the site as a pdf of size: 726 KB

An example of an AACSB Standard is 'Standard 8: Impact of Scholarship". This standard is formally described as :

> 8.1 The school's faculty collectively produce high-quality, impactful intellectual contributions that, over time, develop into mission-consistent areas of thought leadership for the school.
>
> 8.2 The school collaborates with a wide variety of external stakeholders to create and transfer credible, relevant, and timely knowledge that informs the theory, policy, and/or practice of business to develop into mission-consistent areas of thought leadership for the school.
>
> 8.3 The school's portfolio of intellectual contributions contains exemplars of basic, applied, and/or pedagogical research that have had a positive societal impact, consistent with the school's mission.

The standard documentation goes on to define terms within formal description (ie. "Society in this context refers to external stakeholders of relevance to the business school given its mission...").

The standard documentation also outlines the peer review team's 'basis for judgment', and suggests the supporting documentation needed for evaluation of the standard.

For Standard 8, one example of supporting documentation is a five year aggregate portfolio table, which summarizes the publications of faculty.

We curated a collection of 17 accreditation documents from 15 institutions. The institutional data contained accreditation reports and appendices from schools of business across the US of varying program sizes, ranging in date from 2016 to 2023. All documents are publicly available on each institution's website. We use the institution data samples in the 'LLM Based Knowledge Graph Construction' process, defined later in this document.

Models:

There were several considerations taken into account when selecting the models for this project. The first and primary consideration being the suitability of the model for the task. Secondary considerations include performance on the task, accessibility of the model and cost.

This project primarily uses OpenAI's gpt-3.5 models. While this is not the most powerful or recent model available from OpenAI, gpt-3.5 performed well enough on the tasks compared to the gpt-4 model and cost less[3]. This project uses gpt-3.5 for a variety of tasks including document summary classification, knowledge graph building, function calling, multi query and subquery generation, and context-based response generation. With the exception of the multi query generation task where model temperature is set to 0.15 to allow for some creativity, all other model temperatures were set to zero.

GPT-4 is used for the task of Cypher query generation from natural language.

OpenAI's 'text-embedding-ada-002', is used for the task of generating vector embeddings for document text, and input queries.

This project also uses the 'pszemraj/led-large-book-summary'[4] model sourced from HuggingFace for the task of document summarization. For this task we experimented with a more popular (by download count) alternative HuggingFace model 'facebook/bart-large-cnn'[5], but found that the selected model performed better on the task of summarizing our sample business documents.

Please refer to Appendix A , for the model table (A - Table 1) containing all the models explored during development of this project, the appendix also contains a full data source table (A - Table 2).

**Establishing the Knowledge Base**

In this section we discuss each step of establishing our knowledge base; seeding the vector database and constructing the knowledge graph. The process includes both manual data processing and knowledge graph building from pre-processed structured data, and LLM generated knowledge graph data extracted from raw document data. Once the knowledge graph is established we discuss the creation of the semantic layer and vector index.

Manually Building the Knowledge Graph:

*AACSB Standards as Structured Data:*
There are nine AACSB Accreditation Standards and each standard belongs to one of three overall areas. In the standard documentation [26] each standard contains a formal description and three distinct sub-sections: the definitions of terms in the standard, the basis for evaluating the standard and the supporting documentation. The structure of the documentation provided a reasonable starting point for the structure of the hierarchical ontology of the knowledge graph. This hierarchy is illustrated in Figure 3.

---

[3] https://openai.com/pricing

[4] https://huggingface.co/pszemraj/led-large-book-summary
[5] https://huggingface.co/facebook/bart-large-cnn

*Manual Processing:*

We first extracted the "Standard" and "Section" text from the pdf, and organized them into CSV files with the following column headers for the sections: Section_Num, Section_Title, Section_Description . And the following column header names for standards:Section, Standard_num, Standard_title, Standard_formal, Definitions, Basis_for_judement, Supporting_docs.We then performed standard text preprocessing on the imported CSV(s), including standardizing the case of the text, removing English stop words and removing special characters and punctuation – with the exception of periods and newlines. We maintained periods and newlines in consideration of the downstream task of chunking. We also maintained periods due to the format of the text where some descriptions and definitions utilize decimal outline numbering (eg. "1.1") to organize the content, so removing the period from "1.1"would turn it into "11", obscuring the structure and coherence of the text. The processed data was then written to JSON, to provide the structured data for the knowledge graph. This process is illustrated in Figure 3.

This process is available in the repository for this project in the Jupyter Notebook, 'standards_import.ipynb' in the 'notebooks' directory.

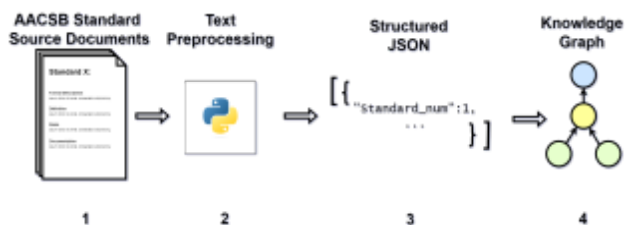

Figure 3. AACSB Standard documentation process steps: 1. Source documentation text extracted from the PDF. 2. Using a Python script, text is preprocessed to standardize the case, remove stop words and remove special characters. 3. Output of the script is a structured JSON file. 4. JSON structured data is used to create the knowledge graph.

*Creating the Knowledge Graph Database:*

To create our knowledge graph database in Neo4j[6], we create node entity structures for all of the standard information in the relevant JSON. As noted above, each standard contains four components. We have organized the components into child nodes that point to their appropriate standard parent node. Each standard child node: "Formal","Basis","Definitions","Documentation" , contains their own child nodes called "Chunks". The phases of the four phases of the node construction are illustrated in Figure 4.

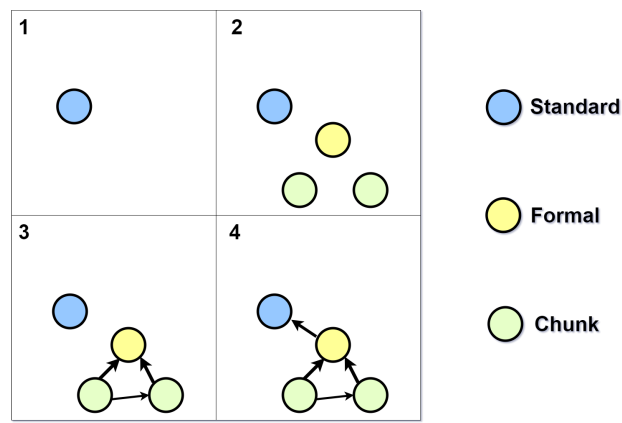

Figure 4. Phases of node construction. For illustrative purposes the diagram is limited to one Standard node with one child node (Formal). Phase 1: Standard nodes are created. Phase 2: Formal node(s) are created along with their associated Chunk nodes. Phase 3. Chunk nodes are linked to each other maintaining the order from the original document, and Chunk nodes are linked to their appropriate parent (Formal). Phase 4. Formal node is linked to the parent Standard.

Chunk nodes contain the text content for their relevant parents. Chunk nodes are structured as an ordered link list to maintain the document structure. The parent standard node is the child node of one of three broader AACSB sections, and the sections belong to the AACSB root node, see Figure 5.

LLM Constructed Knowledge Graph:

*Institutional Documents as Unstructured Data*

The institutional documents provided to the system are unstructured. Given our use-case, institutional users can provide any documentation they see fit including previous accreditation reports, committee meeting notes, department memos, news articles, etc. To create a knowledge graph from this wide variety of input documents we utilize LLMs to generate a knowledge graph from the provided documents.

---

[6] https://neo4j.com/product/neo4j-graph-database/

*LLM-Augmented Knowledge Graph Construction:*
Earlier in this report we detailed the "LLM-Augmented Graph Construction" approach . To implement this in our project we utilize OpenAI function calling [27] to create knowledge graph structured outputs using the 'gpt-3.5-turbo-16k' model. The knowledge graph construction tasks of Entity Typing (ET), Entity Resolution (ER), Coreference Resolution (CR) and Relation Extraction (RE) are handled through prompt engineering. The system prompt provides an a predefined domain-specific list[7] of acceptable node labels, and allows the model to determine the relationships for itself.The prompt also contains instructions on maintaining consistency, handling numerical data and strict compliance rules. The model has the freedom to determine the property types within the node but must include a `parentDocumentUUID`, which is passed into the template function.

Once the unstructured data is processed by the LLM into the Knowledge Graph object, a Graph Document object is created by mapping the node and relationship data to the Graph Document parameters. The resulting graph document is then loaded into the database.

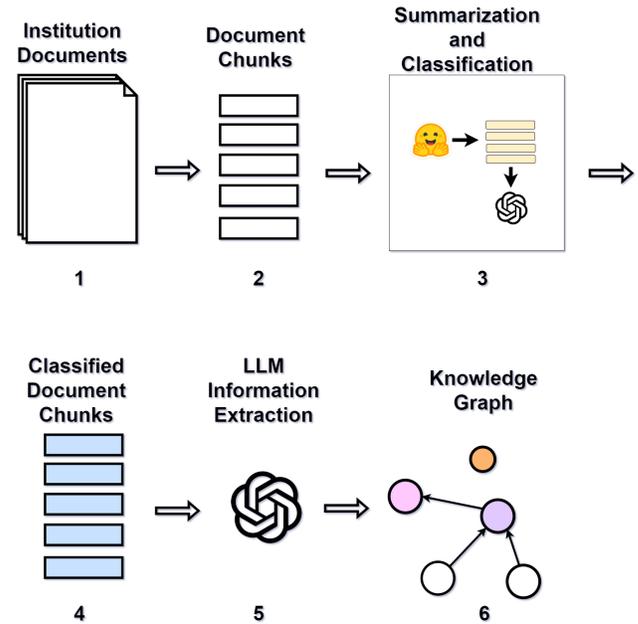

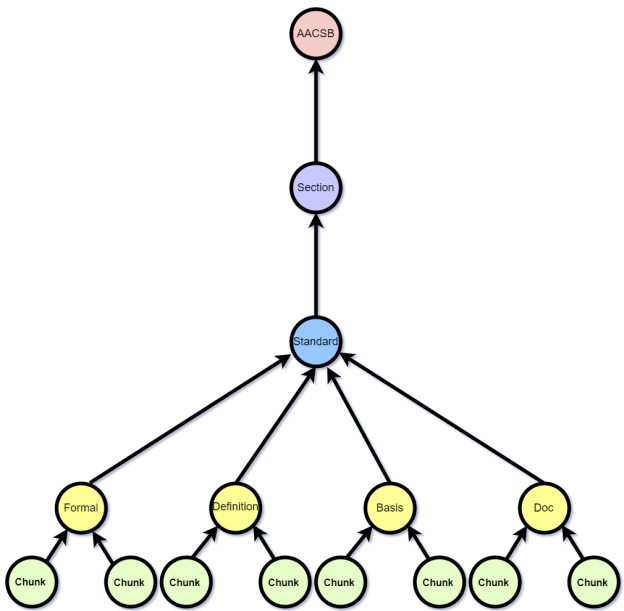

Figure 5. Hierarchy of AACSB node structure. The children of the 'Standard' node are the : Formal (Formal Description), Definition, Basis (Basis for Judgement) and Doc (Documentation) nodes. Each of these nodes contains the 'Chunk' node type, which contains text and the embeddings generated for the semantic layer. Each 'Standard' node belongs to a 'Section' node, and the Section nodes belong to the root AACSB node.

Figure 6. Institution document ingestion process. 1. Institution documents are uploaded. 2.Documents chunked. 3. Document chunks are summarized and classified according to the corresponding AACSB standard. 4. Output classified chunks, containing classification in metadata. 5. Chunks passed in LLM for knowledge graph information extraction (nodes,links). 6. Nodes and links integrated into the knowledge graph.

overview of the task and a persona pattering for the model (ie. 'You are a top-tier algorithm designed for extracting information in structured formats to build a knowledge graph…'). The system prompt contains detailed instructions on labeling nodes according to

This entire process is applied to each chunk of document text provided by the institution. The full prompt and overall LLM to Knowledge Graph pipeline is available in the repository for this project in the Jupyter Notebook, 'llm_to_kg.pynb' in the 'notebooks' directory. Prompts used to extract knowledge graph information from documents can be found in Appendix B.

---

[7] The domain specific list was developed by applying TF-IDF, to hundreds of pages of publicly available accreditation document from several institutions. This approach was used to identify common keywords that were likely to accurately capture node types in the knowledge graph.

*Manual Processing:*
While the LLM handles the knowledge graph construction as it relates to the contents of the institutional documents there is still some manual processing involved to maintain the structure and organization of the source documents. Each 'Document' node is the child of the 'Docsource' node, where the 'Docsource' node contains the import date, and source path of the imported data, there is only one Docsource node per document source import. The child document nodes are linked to each other in a linked list structure to maintain their order, similar to the AACSB structure detailed above. Each document node has the Chunk node as its child. The chunk contains the text and text embedding. Each document is also the parent node to the knowledge graph generated by the LLM. The hierarchy is illustrated in Figure 7. The purpose of this is to link the derived knowledge graph to the structure of the original input source. Ultimately the 'Docsource' nodes are the child nodes of the 'Institution' root node.

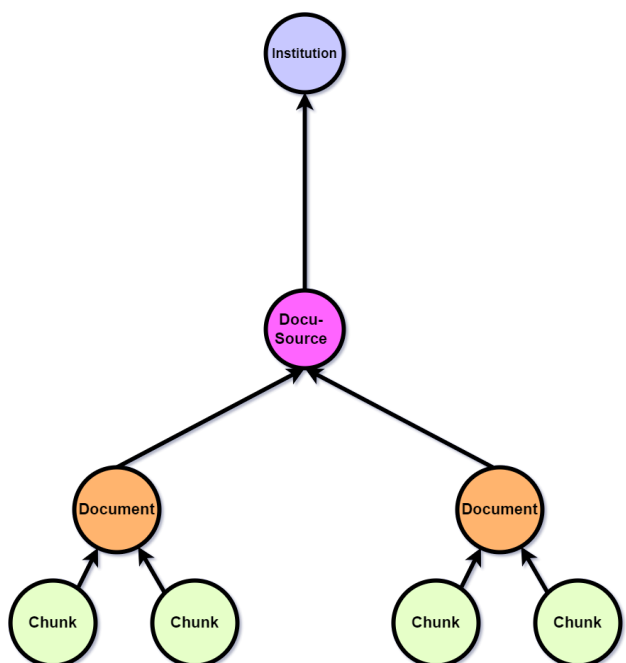

Figure 7. Hierarchy of institution document nodes. Each 'Document' node is related to a 'DocuSource' parent node. Multiple documents can be linked to a single 'DocuSource'. Each 'Document' node has 'Chunk' child nodes which contain the actual text content, and text embedding. The 'Document' node contains the summary of its text content in the 'modelSummary' attribute, and contains the generated AACSB Standard Classification in the 'standardClassification' attribute.

**Bridging the Graphs**

At this point we have detailed two distinct graphs, where the AACSB material has an AACSB root node, and the INSTITUTION material has an INSTITUTION root node. To connect the two graphs we build a link between the institution's 'Document' node and the AACSB standard to which it is most closely related. This is a classification task, where the document is classified according to the standard number (1-9) or 0 for general institutional information. The classification for the document is an integer that is stored in the 'standardClassification' attribute within the 'Document' node.

In reality, it is plausible for a document to be related to multiple AACSB standards. However, in this pipeline design we require the model to output a single classification.

The first step in this process is text summarization. For this task we sourced the 'pszemraj/led-large-book-summary' model from HuggingFace to summarize the text chunk from the document. The model parameters are set to output a summarization with a max_length of 256. Summarizing our input text ensures that in the next step, we stay well within the bounds of the LLM context window in the classification step. The summary is stored in the 'modSummary' attribute of the 'Document' node.

The classification task is handled by `gpt-3.5-turbo-16k` . We pass the summary into a lengthy prompt (appx. 1600 tokens) containing instructions on how to classify the input text as an AACSB standard according to its content. The prompt contains the formal definition of each standard and instructs the model to return the integer of the appropriate standards, or return zero if the input text does not relate to the standards specified in the prompt. The model is instructed to only return an integer with no additional context information, however the model does not always comply. We use a regex pattern to extract the integer from the model output and use that as the classification. This process is illustrated in Figure 6, in step 3, and in Figure 8. As noted above, the final classification is stored in the 'Document' node. And a query is used to build the link between the 'Standard' and the 'Document'.

In Figure 8, we illustrate the linking process with a use case scenario. In our scenario, the institution has added a news article to the system. The news article is about a student entrepreneurship and innovation event. As noted above, the article would be uploaded into the system, summarized by the HuggingFace model and classified by the LLM as being associated with 'Standard 6: Learner Progression'. This standard relates to career development support, among other topics.

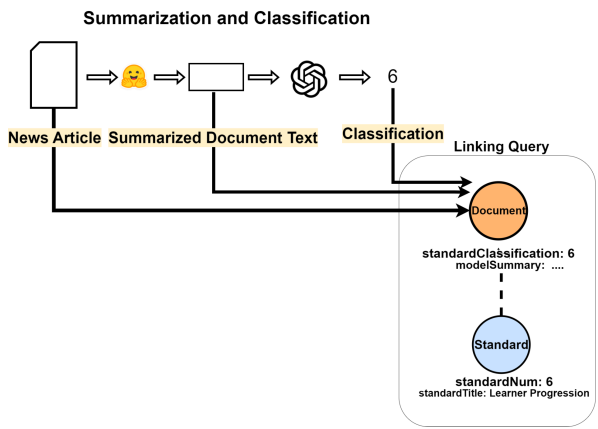

Figure 8. Linking the institution and AACSB knowledge graphs. In this flow a news article on a student entrepreneurship event is passed into the HuggingFace Summarization model. The summarization output is stored in the news article's 'Document' node, and passed into the LLM to be classified according to the related AACSB standard. The LLM classifies the summary, in this case the article summary is classified as 6, for 'Standard 6: Learner Progression'. The classification is stored in the 'standardClassification' attribute of the 'Document' node. Finally, a query is called to link 'Document' nodes with 'Standard' nodes with matching 'standardClassification' and 'standardNum' respectively. The graphs are now linked.

With this classification now stored in the 'Document' node of the article, a query is called to create the link between the 'Document' node and the 'Standard' node for Standard 6.

The full prompt and additional code for the classification and summarization process can be found in the repo for this project in 'input_processing.py' in the 'utils/' directory. The prompt can also be found in Appendix B.

**The Semantic Layer**

We now have a fully constructed knowledge graph capable of handling Cypher queries based on the existing nodes and relationship links between them. The graph is populated with both AACSB accreditation data which is linked to the institutional data, hierarchically organized to maintain document structures along with the extracted semantic[8] knowledge graph data.

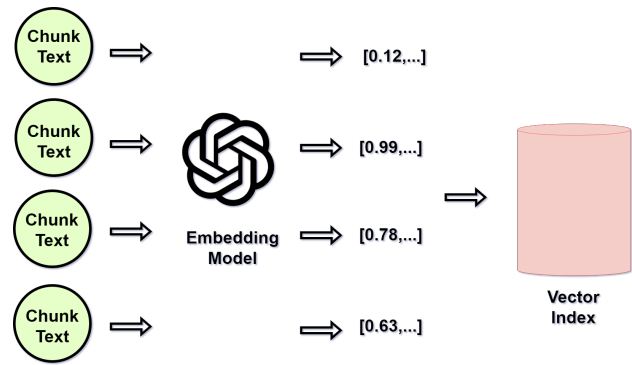

Figure 9. Creating the semantic layer. The 'Chunk' nodes contain all text content for both AACSB Standards and Institution data. To create the embedding layer, the 'Chunk' text is passed into the embedding model, and the created vectors are stored in the vector index.

We introduce a 'semantic layer' to our database, by creating vector embeddings of the text contained in each of the 'Chunk' nodes in both the AACSB and institutional data categories. The embedding task is handled by the 'text-embedding-ada-002' model. The generated embeddings are stored in the 'Chunk' node's 'embedding' attribute, and organized in Neo4J's semantic vector index, see Figure 9. The vector index in this project is called 'accreditation_index' and configured for the cosine similarity metric.

---

[8] 'Semantic' is an overloaded term in this report. 'Semantic' as it relates to knowledge graphs (semantic network), is referring to the meaning extracted from nodes and relationships. 'Semantic' as it refers to vector embeddings relates to the semantic representation of text content.

With the introduction of the semantic layer and the vector index, we can now perform semantic searches on our data, where the queries are embedded into the semantic space and the top k most similar embedding are returned based on the cosine similarity to the query vector.

*Neo4j Vector Index*
Neo4j utilizes Apache Lucene for their vector indexes [28]. Lucene implements a Hierarchical Navigable Small World (HNSW)[29], which approximates the nearest neighbors of an embedding query instead of returning the absolute nearest neighbors based on the similarity metric. Overall this approach provides a more efficient way to search a large dataset in high dimensional spaces, compared to the O(n) operation of searching every entry and finding the exact nearest neighbor.

**Hybrid Context RAG Pipeline**

In this section we review the process for developing our pipeline. The pipeline implements query expansion and query transformation optimization techniques through use of Open AI's function calling for 'multi-query' and 'subquery' generation tasks. Grounding contexts are retrieved from both knowledge graph and vector index, and passed into the generator resulting in the response to the original input query.

*Query Optimization Approaches*
In this project we implement query routing and query expansion to optimize user queries. We use OpenAI Function Calling [30] to handle query routing. The user's original query is first passed into the chat completion to determine if the query is seeking information regarding AACSB information or institution information exclusively, or if the query is seeking information regarding both topic areas.

The chat completion then returns the appropriate function signature to handle the query. We extract the function signature and execute the function based on one of the two scenarios outlined below, and dynamically execute the function call:

Case A: Topic Specific Query
For queries seeking topic specific information, the query is routed to a topic based multi query function, where the query is expanded into multiple topic related queries using a topic specific multi query generation prompt. In this case there are two functions 'generate_multiquestion_institution' and 'generate_multiquestion_aacsb', which return a list of three the generated queries along with other information.

Case B: Multi-Topic Query
For queries encompassing multiple topics, the query is broken down into sub-queries, to identify related topic specific queries. In this case, there is one function 'generate_subquestions_hybrid', which returns a list of two sub queries along with other information. One AACSB related query and one institution related query.

*Multi-Source Retrieval*
The original and generated queries are used to retrieve data from both the knowledge graph and the vector store. To access information from the knowledge graph, the original query is transformed into a Cypher query and executed. Each of the generated queries are converted into vector embeddings where each query is executed on our vector index ,'accreditation_index'. These queries return the top k (k=2) results based on the cosine similarity to the query vector. All query results are collected in a list and converted into a string, which is used as context in the downstream task of generation.

*Generator*
For the task of generating the response to the input query, we use the LLM model 'gpt-3.5-turbo-16k'. We pass the original user query, the context string from the retrieval step and an instruction prompt into the model. The prompt instructs the model to only utilize the context provided and the user query to generate a professionally toned response to the original input query.

The full prompts and additional pipeline code for the retrieval and generation process can be found in the repo for this project in 'retrieverQA.ipynb' in the 'notebooks/' directory. The prompts can also be found in Appendix B**.**

## Results: Hybrid Context RAG Pipeline Evaluation

To evaluate the pipeline we use the Retrieval Augmented Generation Assessment (RAGAs) [31] evaluation framework. To evaluate the pipeline we create a validation set which contains 10 manually created original user queries, ground truth answers, as well as the the context and the answer generated by the pipeline.RAGAs uses each of these fields to calculate the pipeline metrics. The pipeline was evaluated based on faithfulness, answer relevance, context recall and answer relevancy, context recall and answer correctness. These metrics are described below.

### Faithfulness[9]
The faithfulness metric is scaled [0,1], where 1 is the optimal outcome. The faithfulness score is a fraction, where the numerator is the 'Number of claims in the generated answer that can be inferred from a given context' over the "Total number of claims in the generated answer". The metric evaluates the "'actual consistency of the generated answer against the given context".

### Answer Relevancy[10]
Answer relevancy uses the pipeline response (answer), to create LLM-generated synthetic queries. Then, calculates the mean cosine similarity of actual query and the synthetic queries. The relevance score in practice is [0,1], where 1 is the optimal outcome. However, the documentation notes that the range is

not guaranteed given that a cosine similarity score can range [-1,1]

### Context Relevance[11]
Context relevance evaluates the retriever, and measures how well the retrieved context aligns with the question. The context recall is scaled [0, 1] where 1 is the optimal outcome. The metric compares the cardinality of set of sentences "|s|" over the cardinality of the "Total number of sentences in retrieved context".

### Context Recall[12]
Context recall evaluates the retriever, and measures how well the retrieved context aligns with the ground truth. The context recall is scaled [0, 1] where 1 is the optimal outcome. The context recall is a fraction, where the numerator is the cardinality of 'Ground Truth sentences that can be attributed to a context' over the cardinality of "Number of sentences in Ground Truth". The metric evaluates the extent to which "the retrieved context aligns with the annotated answer, treated as the ground truth".

### Answer Correctness[13]
The answer correctness metric is a score ranging [0, 1] where 1 is the optimal outcome. This gauges the accuracy of the pipeline generated response by comparing the ground truth and the generated answer. The calculated metric is a weighted sum between the 'factual correctness' (F1 Score) and the cosine similarity of the answer and truth vectors.

| Table 1: Hybrid Context  RAG Pipeline Metric Summary | | | | | |
|--------|--------------------|--------------|---------------------|-----------------|----------------------|
| Metric | Context Relevance | Faithfulness | Answer Relevancy | Context Recall | Answer Correctness |
| Mean | 0.440 | 0.252 | 0.778 | 0.708 | 0.787 |
| Median | 0.429 | 0.000 | 1.000 | 0.901 | 1.000 |
| Min | 0.000 | 0.000 | 0.000 | 0.000 | 0.000 |
| Max | 0.873 | 1.000 | 1.000 | 0.984 | 1.000 |

Table 1, provides a summary of the pipeline's performance. The average scores across each metric

range from low ("Context Relevance" and "Faithfulness" < 0.45), to fair (Answer Relevancy", "Context Recall" and "Answer Correctness" > 0.70).

The comprehensive metric report can be found in Appendix C. However, we observe that the pipeline on average performed better when evaluating by category of query. The AACSB classified queries pipeline performance on both retrieval and generator tasks outperformed both of the other categories (HYBRID, INST), see Table 5.

| Table 5: Hybrid Context RAG Pipeline Metric Summary (MEAN by category) | | | | | | |
|---|---|---|---|---|---|---|
| Category | | context_relevancy | faithfulness | answer_relevancy | context_recall | answer_correctness |
| AACSB | | 0.524 | 0.522 | 1.000 | 0.900 | 0.813 |
| HYBRID | | 0.217 | 0.000 | 1.000 | 0.915 | 1.000 |
| INST | | 0.481 | 0.000 | 0.333 | 0.300 | 0.611 |

Expanding the size and diversity of questions for the validation query set will provide expanded precision in evaluating the overall pipeline. Given the manual nature of developing the queries, and ground truth information, the process of expanding the dataset is time consuming.

**Observations, Challenges and Future Work:**

At the conclusion of this project we have implemented a vector index and knowledge graph based hybrid context RAG pipeline that integrates large language models into multiple steps of development.

*Observations*
As noted in the results section, pipeline performance on AACSB related queries, outperformed institution related queries achieving optimal scores in 'answer relevancy' and near optimal score in 'context recall'. The performance can likely be attributed to the structured nature of the knowledge graph built manually. Given the consistency of the schema relating to AACSB standards, we are able to more accurately provide references to produce effective and complex Cypher queries. Precise Cypher queries provide additional relevant context from the knowledge graph.

We also observed a notable difference in performance on the document summary classification task. We tested two models for the task of classifying document summaries into 'Standard' categories: 'gpt-3.5-turbo-16k' and 'gpt-4-turbo-preview' both where temperature = 0.

We observed that when using the GPT-4 model, the majority of documents were classified as '0'. Representing a lack of alignment with a specific standard, and categorized as a general institution document.

The poor performance of the GPT-4 model on the classification task can likely be attributed to the model's observed [34] degrading performance on 'composite instruction' prompts. The prompt (Appendix B, Table 1) for our task is a composite prompt, containing several instructions regarding the format of the response to be returned, and how to classify an input based on the standard information provided within the prompt. The issue of GPT-4 either ignoring prompt instructions[14] [15], or producing an inaccurate result [34], is an ongoing issue.

*Challenges*
The primary challenge in working with LLMs, specifically in the task of knowledge graph generation is the non-deterministic information extraction used to construct the knowledge graph. While using the LLM allows for knowledge graphs to be generated from a wide range of unstructured data from institutions, the actual structure of the graph for each use case depends on the output of the LLM at the time of processing. Even when passing the updating the schema into the Cypher Query Template (Appendix B) as a reference, this uncertainty makes generating useful Cypher queries from natural language difficult, as the provided guiding examples in the prompt may not cover the actual schema. For the best results, institutions will need to fine-tune their Cypher query generation for the specific institutional needs.The list of allowed node types will continuously need to be refined, for institution specific types.

---

[14]
https://community.openai.com/t/gpt-4-ignoring-instructions-in-system-prompt/200972
[15]
https://community.openai.com/t/gpt4-turbo-doesnt-listen-to-instructions/588502

The multi-source retrieval approach integrating both vector index data and knowledge graph data provides an additional layer of grounding and context for the generating task. The knowledge graph step provides transparency in the knowledge extraction process, where the vector embedding step requires trust in the embedding model. The hybrid context approach provides a dually grounded context for the generator to produce accurate and relevant responses.

*Future Work*
We note above that institution documents are classified as relating to a single AACSB standard. In the future a multi-label classification implementation could be implemented, either using the LLM approach or a classical machine learning approach on a fine-tined model.

The process creates a complex graph. Additional future work could include experimenting with pruning the LLM constructed knowledge graph to remove superfluous links while maintaining connectedness and accuracy.

Overall the project produces a viable hybrid context pipeline that can continue to be refined and improved.

[1] E. A. Vogels, "A majority of Americans have heard of CHATGPT, but few have tried it themselves," Pew Research Center, https://www.pewresearch.org/short-reads/2023/05/24/a-majority-of-americans-have-heard-of-chatgpt-but-few-have-tried-it-themselves/ (accessed Apr. 18, 2024).

[2] T. Kojima, S. S. Gu, M. Reid, Y. Matsuo, and Y. Iwasawa, "Large language models are zero-shot Reasoners," arXiv.org, https://arxiv.org/abs/2205.11916 (accessed Apr. 18, 2024).

[3] W. X. Zhao et al., "A survey of large language models," arXiv.org, https://arxiv.org/abs/2303.18223 (accessed Apr. 22, 2024).

[4] H. Naveed et al., "A comprehensive overview of large language models," arXiv.org, https://arxiv.org/abs/2307.06435 (accessed Apr. 22, 2024).

[5] Z. Xu, S. Jain, and M. Kankanhalli, "Hallucination is inevitable: An innate limitation of large language models," arXiv.org, https://arxiv.org/abs/2401.11817 (accessed Apr. 22, 2024).

[6] P. Lewis et al., "Retrieval-augmented generation for knowledge-intensive NLP tasks," arXiv.org, https://arxiv.org/abs/2005.11401 (accessed Apr. 22, 2024).

[7] Y. Gao et al., "Retrieval-augmented generation for large language models: A survey," arXiv.org, https://arxiv.org/abs/2312.10997 (accessed Apr. 22, 2024).

[8] W. Yu et al., "Generate rather than retrieve: Large language models are strong context generators," arXiv.org, https://arxiv.org/abs/2209.10063 (accessed Apr. 22, 2024).

[9] O. Khattab et al., "Demonstrate-search-predict: Composing retrieval and language models for knowledge-intensive NLP," arXiv.org, https://arxiv.org/abs/2212.14024 (accessed Apr. 22, 2024).

[10] R. Schwaber-Cohen, "Chunking Strategies for LLM Applications," Pinecone, https://www.pinecone.io/learn/chunking-strategies/ (accessed Apr. 22, 2024).

[11] "Semantic chunking," 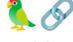 LangChain, https://python.langchain.com/docs/modules/data_connection/document_transformers/semantic-chunker/ (accessed Apr. 22, 2024).

[12] "Semantic chunker¶," LlamaIndex, https://docs.llamaindex.ai/en/stable/examples/node_parsers/semantic_chunking/ (accessed Apr. 22, 2024).

[13] Y. Wang et al., "Knowledge graph prompting for multi-document question answering," arXiv.org, https://arxiv.org/abs/2308.11730 (accessed Apr. 22, 2024).

[14] Z. Rackauckas, "Rag-fusion: A new take on retrieval-augmented generation," arXiv.org, https://arxiv.org/abs/2402.03367 (accessed Apr. 22, 2024).

[15] S. Dhuliawala et al., "Chain-of-verification reduces hallucination in large language models," arXiv.org, https://arxiv.org/abs/2309.11495 (accessed Apr. 22, 2024).

[16] "What is a knowledge graph?," IBM, https://www.ibm.com/topics/knowledge-graph (accessed Apr. 22, 2024).

[17] B. Abu-Salih and S. Alotaibi, "A systematic literature review of Knowledge Graph Construction and application in education," Heliyon, https://www.ncbi.nlm.nih.gov/pmc/articles/PMC10847940/ (accessed Apr. 22, 2024).

[18] A. Singhal, "Introducing the knowledge graph: Things, not strings," Google, https://blog.google/products/search/introducing-knowledge-graph-things-not/ (accessed Apr. 22, 2024).

[19] M. Kejriwal, "Knowledge graphs: A practical review of the research landscape," MDPI, https://www.mdpi.com/2078-2489/13/4/161 (accessed Apr. 22, 2024).

[20] I. Tiddi, "Knowledge graphs as tools for explainable machine learning: A survey," Artificial Intelligence, https://www.sciencedirect.com/science/article/pii/S0004370221001788 (accessed Apr. 22, 2024).

[21] M. Hofer, D. Obraczka, A. Saeedi, H. Köpcke, and E. Rahm, "Construction of knowledge graphs: State and challenges," arXiv.org, https://arxiv.org/abs/2302.11509 (accessed Apr. 22, 2024).

[22] L. Zhong, J. Wu, Q. Li, H. Peng, and X. Wu, "A comprehensive survey on Automatic Knowledge Graph Construction," arXiv.org, https://arxiv.org/abs/2302.05019 (accessed Apr. 22, 2024).

[23] L. Yang, H. Chen, Z. Li, X. Ding, and X. Wu, "Give us the facts: Enhancing large language models with knowledge graphs for fact-aware language modeling," arXiv.org, https://arxiv.org/abs/2306.11489 (accessed Apr. 22, 2024).

[24] S. Pan et al., "Unifying large language models and knowledge graphs: A roadmap," arXiv.org, https://arxiv.org/abs/2306.08302 (accessed Apr. 22, 2024).

[25] B. Jehangir "A survey on named entity recognition - datasets, tools, and methodologies," Natural Language Processing Journal, https://www.sciencedirect.com/science/article/pii/S2949719123000146 (accessed Apr. 22, 2024).

[26] "AACSB Business Accreditation Standards," AACSB, https://www.aacsb.edu/educators/accreditation/business-accreditation/aacsb-business-accreditation-standards (accessed Apr. 22, 2024).

[27] Function calling, https://platform.openai.com/docs/guides/function-calling (accessed Apr. 23, 2024).

[28] "Vector search indexes - cypher manual," Neo4j Graph Data Platform, https://neo4j.com/docs/cypher-manual/current/indexes/semantic-indexes/vector-indexes/ (accessed Apr. 22, 2024).

[29] Yu. A. Malkov and D. A. Yashunin, "Efficient and robust approximate nearest neighbor search using hierarchical navigable small world graphs," arXiv.org, https://arxiv.org/abs/1603.09320 (accessed Apr. 22, 2024).

[30] Function calling, https://platform.openai.com/docs/guides/function-calling (accessed Apr. 23, 2024).

[31] S. Es, J. James, L. Espinosa-Anke, and S. Schockaert, "Ragas: Automated Evaluation of retrieval augmented generation," arXiv.org, https://arxiv.org/abs/2309.15217 (accessed Apr. 22, 2024).

[32] LangChain, "Constructing knowledge graphs from text using openai functions: Leveraging knowledge graphs to Power Langchain applications," LangChain Blog, https://blog.langchain.dev/constructing-knowledge-graphs-from-text-using-openai-functions/ (accessed Apr. 22, 2024).

[33] "NEO4J Cypher Template,"   LangChain, https://python.langchain.com/docs/integrations/graphs/neo4j_cypher/ (accessed Apr. 22, 2024).

[34] L. Chen, M. Zaharia, and J. Zou, "How is CHATGPT's behavior changing over time?," arXiv.org, https://arxiv.org/abs/2307.09009 (accessed Apr. 24, 2024).

[INTENTIONALLY LEFT BLANK]

**APPENDICES:**
A: Data and Models
B: Prompt Engineering
C: Pipeline Evaluation
D: System Flow and Graph  Diagrams

APPENDIX A:  DATA AND MODELS

| Table 1: Project Model Summary | | | | | |
|---|---|---|---|---|---|
| **Name** | **Project Task** | **Used/Live In Project** | **Provider** | **Input: Cost per 1k tokens** | **Output: Cost per 1k tokens** |
| gpt-3.5-turbo-0125 | function calling | TRUE | OpenAI | 0.0005 | 0.0015 |
| gpt-3.5-turbo-16k | generator | TRUE | OpenAI | 0.0005 | 0.0015 |
| | document summary classification | TRUE | OpenAI | | |
| | knowledge graph building | TRUE | OpenAI | | |
| | multi-query/sub-query question development | TRUE | OpenAI | | |
| | cypher retrieval qa | TRUE | OpenAI | | |
| gpt-4 | cypher query generation | TRUE | OpenAI | 0.03 | 0.06 |
| | document summary classification | FALSE | OpenAI | | |
| text-embedding-ada-002 | text embedding | TRUE | OpenAI | 0.0001 (per call - not token) | n/a |
| text-embedding-3-small | text embedding | FALSE | OpenAI | 0.00002 (per call - not token) | n/a |
| pszemraj/led-large-book-summary | summarization | TRUE | HuggingFace | Free | Free |
| facebook/bart-large-cnn | summarization | FALSE | HuggingFace | Free | Free |

| Table 2: Institutional Data Source List | | | | | |
|---|---|---|---|---|---|
| **Institution** | **Business School** | **Report/PDF Source Link** | **Type** | | **YEAR** |
| Texas State University | McCoy College of Business | https://www.mccoy.txst.edu/faculty-staff/faculty-resources/aacsb.html | CIR | | 2016 |
| UT Tyler | Soules College of Business | https://www.uttyler.edu/soules-college-of-business/policies-and-procedures/2023continuous_improvement_report.pdf | CIR | | 2023 |
| Loyola University of New Orleans | College of Business | http://business.loyno.edu/sites/business.loyno.edu/files/AACSB_Report_LoyolaNO_1-20-23.pdf<br>http://business.loyno.edu/aacsb-accreditation | CIR | | 2023 |
| University of North Texas | G. Brint Ryan College of Business | https://cob.unt.edu/pdf/University%20of%20North%20Texas%20College%20of%20Business%20Department%20of%20Accounting%20Fifth%20Year%20Continuous%20Improvement%20Review%20Report?/sites/default/files/docs/dean/UNT%20ACCT%20Complete%20AACSB%20Report.pdf | CIR | ACCOUNTING | |
| Valdosta State University | Langdale College of Business | https://www.valdosta.edu/colleges/business/deans-office/documents/aacsb.pdf | CIR | | 2016 |
| Southeastern Oklahoma State University | John Massey School of Business | https://www.se.edu/business/wp-content/uploads/sites/5/2019/06/JMSB-AACSB-Report-Final-document.pdf | CIR | | 2019 |
| Grambling State University | College of Business | https://www.gram.edu/academics/majors/business/research/docs/Final%202015-%202019%20Fifth%20Year%20AACSB%20Accreditation%20Maintenance%20Report.pdf | CIR | | 2019 |
| Villanova University | Villanova School of Business | https://www1.villanova.edu/dam/villanova/VSB/assets/AACSB/VSB/VSB%20AACSB%20Report%20for%20Fac%20Staff%20Dist.pdf | CIR | | 2017 |
| University of Wisconsin Parkside | College of Business and Economic Computing | https://www.uwp.edu/learn/departments/business/upload/CIR-Report-Final-9-1-2020.pdf | CIR | | 2020 |
| Western Illinois University | College of Business and Technology | https://www.wiu.edu/cbt/documents/CBT%20Business%20Assessment.pdf<br>https://www.wiu.edu/cbt/aacsb.php | CIR | | 2020 |
| Dalton State | Wright School of Business | https://www.daltonstate.edu/skins/userfiles/files/AACSB%20CIR%20WSOB%20FINAL%206-27-2019.pdf | CIR | | 2019 |
| University of San Francisco | School of Management | https://myusf.usfca.edu/sites/default/files/users/camara/USF_SOM_AACSB%20CIR%202020%20FINAL_25November2020_Corrected.pdf | CIR | | 2020 |
| University of Northern Colorado | Monfort College of Business | https://mcb.unco.edu/pdf/aacsb/Monfort-College-of-Business-Self-Study-Report-8-2-17.pdf | CIR | | 2017 |
| University of South Florida | MUMA College of Business | https://www.usf.edu/business/documents/accreditation/business-report-continuous-improvement.pdf | CIR | | 2017 |
| Georgia College and State University | J. Whitney Bunting College of Business and Technology | https://www.gcsu.edu/business/relevant-resources-and-materials-business | CIR | | 2016 |

APPENDIX B: PROMPT ENGINEERING

| Table 1: Prompts by Task | |
|---|---|
| **Task** | **Prompt** |

Document
Summary
Classification

## 3. AACSB Standard Definitions

For more context take into consideration the following descriptions of each standard

#### Standard 1: Strategic Planning

The school maintains a well-documented strategic plan, developed through a robust and collaborative planning process involving key stakeholder input, that informs the school on resource allocation priorities. The strategic plan should also articulate a clear and focused mission for the school.

1.2 The school regularly monitors its progress against its planned strategies and expected outcomes and communicates its progress to key stakeholders. As part of monitoring, the school conducts formal risk analysis and has plans to mitigate identified major risks.

1.3 As the school carries out its mission, it embraces innovation as a key element of continuous improvement.

1.4 The school demonstrates a commitment to positive societal impact as expressed in and supported by its focused mission and specifies how it intends to achieve this impact.

#### Standard 2: Physical, Virtual and Financial Resources

2.1 physical, 2.2 virtual, and 2.3 financial resources to sustain the school on an ongoing basis and to promote a high-quality environment that fosters success of all participants in support of the school's mission, strategies, and expected outcomes.

#### Standard 3:Faculty and Professional Staff Resources

3.1 The school maintains and strategically deploys sufficient participating and supporting faculty who collectively demonstrate significant academic and professional engagement that, in turn, supports high-quality outcomes consistent with the school's mission.

3.2 Faculty are qualified through initial academic or professional preparation and sustain currency and relevancy appropriate to their classification, as follows: Scholarly Academic (SA), Practice Academic (PA), Scholarly Practitioner (SP), or Instructional Practitioner (IP). Otherwise, faculty members are classified as Additional Faculty (A).

3.3 Sufficient professional staff are available to ensure high-quality support for faculty and learners as appropriate.

3.4 The school has well-documented and well-communicated processes to manage, develop, and support faculty and professional staff over the progression of their careers that are consistent with the school's mission, strategies, and expected outcomes.

#### Standard 4: Curriculum

4.1 The school delivers content that is current, relevant, forward-looking, globally oriented, aligned with program competency goals,

and consistent with its mission, strategies, and expected outcomes. The curriculum content cultivates agility with current and emerging technologies.

4.2 The school manages its curriculum through assessment and other systematic review processes to ensure currency, relevancy, and competency.

4.3 The school's curriculum promotes and fosters innovation, experiential learning, and a lifelong learning mindset. Program elements promoting positive societal impact are included within the curriculum.

4.4 The school's curriculum facilitates meaningful learner-to-learner and learner to-faculty academic and professional engagement.

#### Standard 5: Assurance of Learning

5.1 The school uses well-documented assurance of learning (AoL) processes that include direct and indirect measures for ensuring the quality of all degree programs that are deemed in scope for accreditation purposes. The results of the school's AoL work leads to curricular and process improvements.

5.2 Programs resulting in the same degree credential are structured and designed to ensure equivalence of high-quality outcomes irrespective of location and modality of instructional delivery.

5.3 Microlearning credentials that are "stackable" or otherwise able to be combined into an AACSB-accredited degree program should include processes to ensure high quality and continuous improvement.

5.4 Non-degree executive education that generates greater than five percent of a school's total annual resources should include processes to ensure high quality and continuous improvement.

#### Standard 6: Learner Progression

6.1 The school has policies and procedures for admissions, acceptance of transfer credit, academic progression toward degree completion, and support for career development that are clear, effective, consistently applied, and aligned with the school's mission, strategies, and expected outcomes.

6.2 Post-graduation success is consistent with the school's mission, strategies, and expected outcomes. Public disclosure of academic program quality supporting learner progression and post-graduation success occurs on a current and consistent basis.

#### Standard 7: Teaching Effectiveness and Impact

7.1 The school has a systematic, multi-measure assessment process for ensuring quality of teaching and impact on learner success.

7.2 The school has development activities in place to enhance faculty teaching and ensure that teachers can deliver curriculum that is current, relevant, forward looking, globally oriented, innovative, and aligned with program competency goals.

7.3 Faculty are current in their discipline and pedagogical methods, including teaching diverse perspectives in an inclusive environment. Faculty demonstrate a lifelong learning mindset, as supported and promoted by the school.

7.4 The school demonstrates teaching impact through learner success, learner satisfaction, and other affirmations of teaching expertise

#### Standard 8: Impact of Scholarship

8.1 The school's faculty collectively produce high-quality, impactful intellectual contributions that, over time, develop into mission-consistent areas of thought leadership for the school.

8.2 The school collaborates with a wide variety of external stakeholders to create and transfer credible, relevant, and timely knowledge that informs the theory, policy, and/or practice of business to develop into mission-consistent areas of thought leadership for the school.

8.3 The school's portfolio of intellectual contributions contains exemplars of basic, applied, and/or pedagogical research that have had a positive societal impact, consistent with the school's mission.

#### Standard 9: Engagement and Societal Impact

9.1 The school demonstrates positive societal impact through internal and external initiatives and/or activities, consistent with the school's mission, strategies, and expected outcomes.

## 4. Discernment Notes

Standard 4 relates to curriculum, Standard 5 relates to assurance of learning. You must classify text that deals the process of evaluating performance as Standard 5, evaluation often includes percentages, comparison to previous years, sample sizes etc,

ONLY classify text as standard 4 if it strictly discusses curriculum.

## 5. Strict Compliance

Evaluate the input text using only the above standard information. You must be correct in your classification, if any input does not meet the standard classification criteria then classify it as 0 , which is general institution information

YOU MUST ONLY RETURN AN INTEGER NO OTHER ADDITIONAL CONTEXT INFORMATION. Failure to comply will result in termination.

"""

Knowledge Graph Extraction ♣

f"""# Knowledge Graph Instructions for GPT-3.5 Turbo
## 1. Overview
You are a top-tier algorithm designed for extracting information in structured formats to build a knowledge graph. You are evaluating information from an academic institution, creating a knowledge graph of concepts related to AACSB accreditation.
You should be able to identify learning goals (eg. Written Communication, Oral Communication, Critical Thinking, Ethics, Globalization, Information Technology),
along with how and when they are assessed.
- **Nodes** represent entities and concepts. They're akin to Wikipedia nodes.
- The aim is to achieve simplicity and clarity in the knowledge graph, making it accessible for a vast audience.
## 2. Labeling Nodes
- **Consistency**: Ensure you use basic or elementary types for node labels.
- For example, when you identify an entity representing a person, always label it as **"person"**. Avoid using more specific terms like "mathematician" or "scientist".
- **Node IDs**: Never utilize integers as node IDs. Node IDs should be names or human-readable identifiers found in the text.
{'- **Allowed Node Labels:**' + ", ".join(allowed_nodes) if allowed_nodes else ""}
{'- **Allowed Relationship Types**:' + ", ".join(allowed_rels) if allowed_rels else ""}
## 3. Handling Numerical Data and Dates
- Numerical data, like age or other related information, should be incorporated as attributes or properties of the respective nodes.
- **No Separate Nodes for Dates/Numbers**: Do not create separate nodes for dates or numerical values. Always attach them as attributes or properties of nodes.
- ** Provided REQUIRED Property**: Each node must have a the property key "parentDocUUID", this is the value {parentDocUUID}. DO NOT CREATE A VALUE only use the provided value.
- **Property Format**: Properties must be in a key-value format.
- **Quotation Marks**: Never use escaped single or double quotes within property values.
- **Naming Convention**: Use camelCase for property keys, e.g., `birthDate`.
## 4. Coreference Resolution
- **Maintain Entity Consistency**: When extracting entities, it's vital to ensure consistency.
If an entity, such as "John Doe", is mentioned multiple times in the text but is referred to by different names or pronouns (e.g., "Joe", "he"),
always use the most complete identifier for that entity throughout the knowledge graph. In this example, use "John Doe" as the entity ID.
Remember, the knowledge graph should be coherent and easily understandable, so maintaining consistency in entity references is crucial.
## 5. Strict Compliance

| | |
|---|---|
| | Adhere to the rules strictly. Non-compliance will result in termination.<br>""" |
| Retriever:<br>Multi-query<br>AACSB | f"""<br><br># Instruction<br>You are an AI language model assistant. Your task is to generate three<br>different versions of the given user question to retrieve relevant documents from a vector<br>database. By generating multiple perspectives on the user question, your goal is to help<br>the user overcome some of the limitations of the distance-based similarity search, by providing precise<br>questions.<br>Provide these alternative questions separated by "[SEP]" TOKEN.<br><br># List Delimiter<br>Each question in the list must be separated by the "[SEP]" TOKEN<br><br># Important Context: AACSB<br>The questions you generate are directly related to extracting useful information on the<br>AACSB accreditation standards. The questions you generate will be used as vector database index<br>queries that contain information on:<br><br>- formal AACSB descriptions<br>- documentation that supports each standard<br>- basis for evaluation of the standards<br>- relevant definitions of terms used in the standard descriptions.<br><br>The AACSB Website provides the following summary of their work:<br>AACSB accreditation is known, worldwide, as the longest-standing, most recognized form of<br>specialized accreditation that an institution and its business programs can earn.<br>Accreditation is a voluntary, nongovernmental process that includes a rigorous external review<br>of a school's mission, faculty qualifications, curricula, and ability to provide the highest-quality programs.<br><br># Format Rules<br>DO NOT NUMBER THE LIST<br>DO NOT ANSWER THE QUESTION<br>DELIMITER USING "[SEP]" token<br>Original question: {question}<br>""" |
| | f"""<br><br># Instruction<br>You are an AI language model assistant. Your task is to generate three<br>different versions of the given user question to retrieve relevant documents from a vector<br>database. By generating multiple perspectives on the user question, your goal is to help<br>the user overcome some of the limitations of the distance-based similarity search, by providing precise<br>questions.<br>Provide these alternative questions separated by "[SEP]" TOKEN.<br><br># List Delimiter<br>Each question in the list must be seperated by the "[SEP]" TOKEN<br><br># Important Context: Academic Institution<br>The questions you generate are directly related to extracting useful information about a School<br>of Business. The questions you generate will be used as vector database index<br>queries that contain information on:<br><br>- Strategic Plan, Mission and Fiscal Resources<br>- Academic Departments in the School of Business inluding not limited to : Accounting, Marketing, Management, Finance, Entreprenuership<br>- Student Services and Student Organizations<br>- Program Goals, Learning Objectives and Curriculum Assessment<br>- Continuous Improvement<br><br><br># Format Rules<br>DO NOT NUMBER THE LIST<br>DO NOT ANSWER THE QUESTION<br>DELIMITER USING "[SEP]" token<br>Original question: {question}<br>""" |
| Retriever: Multi-query Institution | |
| | f"""<br><br># Instruction<br>You are an AI language model assistant. Your task is to evaluate the users input query and<br>divide the query into two and ONLY TWO sub questions. The first sub questions should address the portion<br>of the user query that relates to the AACSB Standards, the second subquestion should relate the institution<br>specific portion of the query. Your overall objective is to break down the complex user query into the<br>two distinct sub questions. Provide these alternative questions separated by "[SEP]" TOKEN.<br><br># List Delimiter<br>Each question in the list must be separated by the "[SEP]" TOKEN<br><br># Important Context: Sub<br><br>## 1. AACSB Standard sub question:<br>AACSB sub question may related to accreditation content such as:<br><br>- formal AACSB descriptions<br>- documentation that supports each standard<br>- basis for evaluation of the standards<br>- relevant definitions of terms used in the standard descriptions.<br><br>The AACSB Website provides the following summary of their work:<br>AACSB accreditation is known, worldwide, as the longest-standing, most recognized form of<br>specialized accreditation that an institution and its business programs can earn.<br>Accreditation is a voluntary, nongovernmental process that includes a rigorous external review<br>of a school's mission, faculty qualifications, curricula, and ability to provide the highest-quality programs.<br><br><br>## 2. Academic Institution sub question:<br>Academic Institution sub question may relate to School of Business content such as:<br><br>- Strategic Plan, Mission and Fiscal Resources<br>- Academic Departments in the School of Business including not limited to : Accounting, Marketing,<br>Management, Finance, Entrepreneurship<br>- Student Services and Student Organizations<br>- Program Goals, Learning Objectives and Curriculum Assessment<br>- Continuous Improvement<br><br><br># Format Rules<br>DO NOT NUMBER THE LIST<br>DO NOT ANSWER THE QUESTION<br>DELIMITER USING "[SEP]" token<br>Original question: {question}<br>""" |
| Retriever:<br>Hybrid Sub<br>query | |
| | """Task:Generate Cypher statement to query a graph database.<br>Instructions:<br>Use only the provided relationship types and properties in the schema.<br>Do not use any other relationship types or properties that are not provided.<br><br>Schema:<br>{schema}<br>Note: Do not include any explanations or apologies in your responses.<br>Do not respond to any questions that might ask anything else than for you to construct a Cypher statement.<br>Do not include any text except the generated Cypher statement.<br><br>Perfect Syntax: Your queries must be in the correct Cypher syntax, at all costs you should avoide<br>'CypherSyntaxError' and 'ValueError' resulting from your query ie:<br>ValueError: Generated Cypher Statement is not valid<br>code: Neo.ClientError.Statement.SyntaxError message: Invalid input 'objective': expected ")", "WHERE",<br>or a parameter (line 1, column 19 (offset: 18))<br>"MATCH (n:Learning objective) RETURN n.name"<br><br>Examples: Here are a few examples of generated Cypher statements for particular questions:<br><br>## AACSB STANDARD EXAMPLE<br># Which standards deal with staff resources?<br>MATCH (n)<br>WHERE n.nodeCat = 'AACSB' AND (n.text CONTAINS 'staff' AND n.text CONTAINS 'resources' OR<br>n.text CONTAINS 'staff resources')<br>RETURN n<br><br># How is standard 2 documented<br>MATCH (d:Documentation) WHERE d.parentStandardNum = 2 RETURN d.text<br><br>## INSTITUTION EXAMPLE -- schema may vary from example, reference schema<br># Which learning objectives did undergradute and graduate program evaluate<br>MATCH (p:Program)-[]->(l:\`Learning objective\`)<br>WHERE (p.name CONTAINS 'undergraduate' OR p.name CONTAINS 'graduate')<br>RETURN l.name |
| Cypher Query Generator‡ | The question is:<br>{question}""" |

| |
|---|
| ♣: Foundation prompt from [32], then modified/expanded |
| ‡: Foundation prompt from [33], then modified/expanded |

APPENDIX C: PIPELINE EVALUATION

**Table 1: Hybrid Context RAG Pipeline Metric Summary**

| Metric | Context Relevance | Faithfulness | Answer Relevancy | Context Recall | Answer Correctness |
|---|---|---|---|---|---|
| Mean | 0.440 | 0.252 | 0.778 | 0.708 | 0.787 |
| Median | 0.429 | 0.000 | 1.000 | 0.901 | 1.000 |
| Min | 0.000 | 0.000 | 0.000 | 0.000 | 0.000 |
| Max | 0.873 | 1.000 | 1.000 | 0.984 | 1.000 |

**Table 2: Hybrid Context RAG Comprehensive Metric Report (by query)**

| qID | context_relevancy | faithfulness | answer_relevancy | context_recall | answer_correctness |
|---|---|---|---|---|---|
| 1 | 0.000 | 0.000 | 1.000 | 0.780 | 1.000 |
| 2 | 0.217 | 0.000 | 1.000 | 0.915 | 1.000 |
| 3 | 0.873 | 0.143 | 1.000 | 0.937 | 1.000 |
| 4 | 0.793 | 1.000 | 1.000 | 0.984 | 0.750 |
| 5 | 0.429 | 0.944 | 1.000 | 0.900 | 0.500 |
| 6 | 0.471 | 0.000 | 0.000 | 0.000 | 0.000 |
| 7 | 0.194 | 0.000 | 1.000 | 0.901 | 0.833 |
| 9 | 0.201 | 0.176 | 1.000 | 0.959 | 1.000 |
| 10 | 0.777 | 0.000 | 0.000 | 0.000 | 1.000 |

*Note: Excluded query with {qID:8} (Issue)

**Table 3: Pipeline Test Queries List**

| qID | question | Category |
|---|---|---|
| 1 | AACSB Standards on sustainability | AACSB |
| 2 | Our accounting department updated curriculum to include environmental impact in business risk evaluation, does this reflect the sustainability standard | HYB |
| 3 | Standard 5 specifies a systematic process for assurance of learning. What do peer review teams usually expect in determining whether this standard is met?[16] | AACSB |
| 4 | What are intellectual contributions | AACSB |
| 5 | Must faculty members publish in order to be qualified as scholarly academic, practice | AACSB |

---

[16] https://www.aacsb.edu/educators/accreditation/business-accreditation/faqs

| | academic, or scholarly practitioner?[17] | |
|---|---|---|
| 6 | Who is the dean of my school | INST |
| 7 | What are the management courses | INST |
| 8 | Which learning objectives are assessed in the undergraduate business program | INST |
| 9 | How is standard 8 defined? | AACSB |
| 10 | What are MACC program students expected to demonstrate | INST |

**Table 4: Hybrid Context RAG Comprehensive Metric Report (by category)**

| Category | qID | context_relevancy | faithfulness | answer_relevancy | context_recall | answer_correctness |
|---|---|---|---|---|---|---|
| AACSB | 1 | 0.000 | 0.000 | 1.000 | 0.780 | 1.000 |
| | 3 | 0.873 | 0.143 | 1.000 | 0.937 | 1.000 |
| | 4 | 0.793 | 1.000 | 1.000 | 0.984 | 0.750 |
| | 5 | 0.429 | 0.944 | 1.000 | 0.900 | 0.500 |
| HYBRID | 2 | 0.217 | 0.000 | 1.000 | 0.915 | 1.000 |
| INST | 6 | 0.471 | 0.000 | 0.000 | 0.000 | 0.000 |
| | 7 | 0.194 | 0.000 | 1.000 | 0.901 | 0.833 |
| | 10 | 0.777 | 0.000 | 0.000 | 0.000 | 1.000 |

**Table 5: Hybrid Context RAG Pipeline Metric Summary (by category)**

| Category | | context_relevancy | faithfulness | answer_relevancy | context_recall | answer_correctness |
|---|---|---|---|---|---|---|
| AACSB | | 0.524 | 0.522 | 1.000 | 0.900 | 0.813 |
| HYBRID | | 0.217 | 0.000 | 1.000 | 0.915 | 1.000 |
| INST | | 0.481 | 0.000 | 0.333 | 0.300 | 0.611 |

---

[17] https://www.aacsb.edu/educators/accreditation/business-accreditation/faqs/faculty-qualifications

APPENDIX D: SYSTEM FLOW DIAGRAMS AND GRAPH SNAPSHOTS

Figure 1: Hybrid Context RAG Pipeline

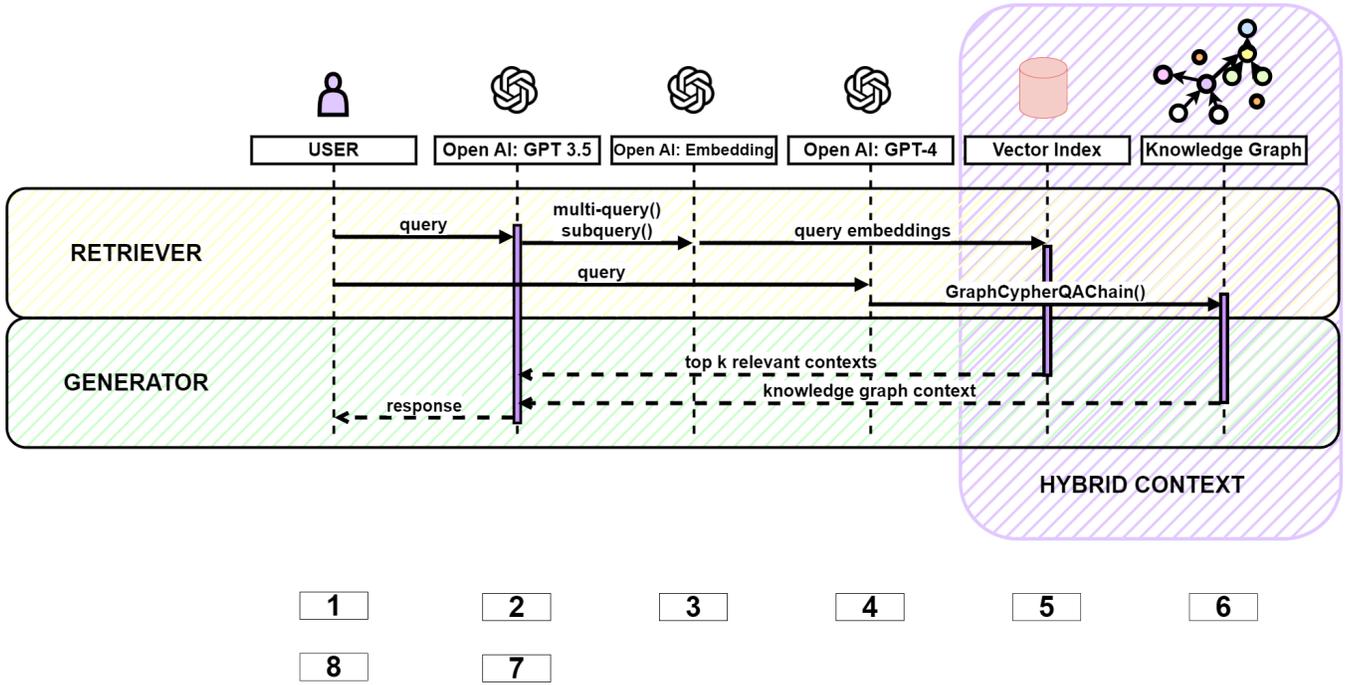

Figure 2: Data Ingest Flow Process

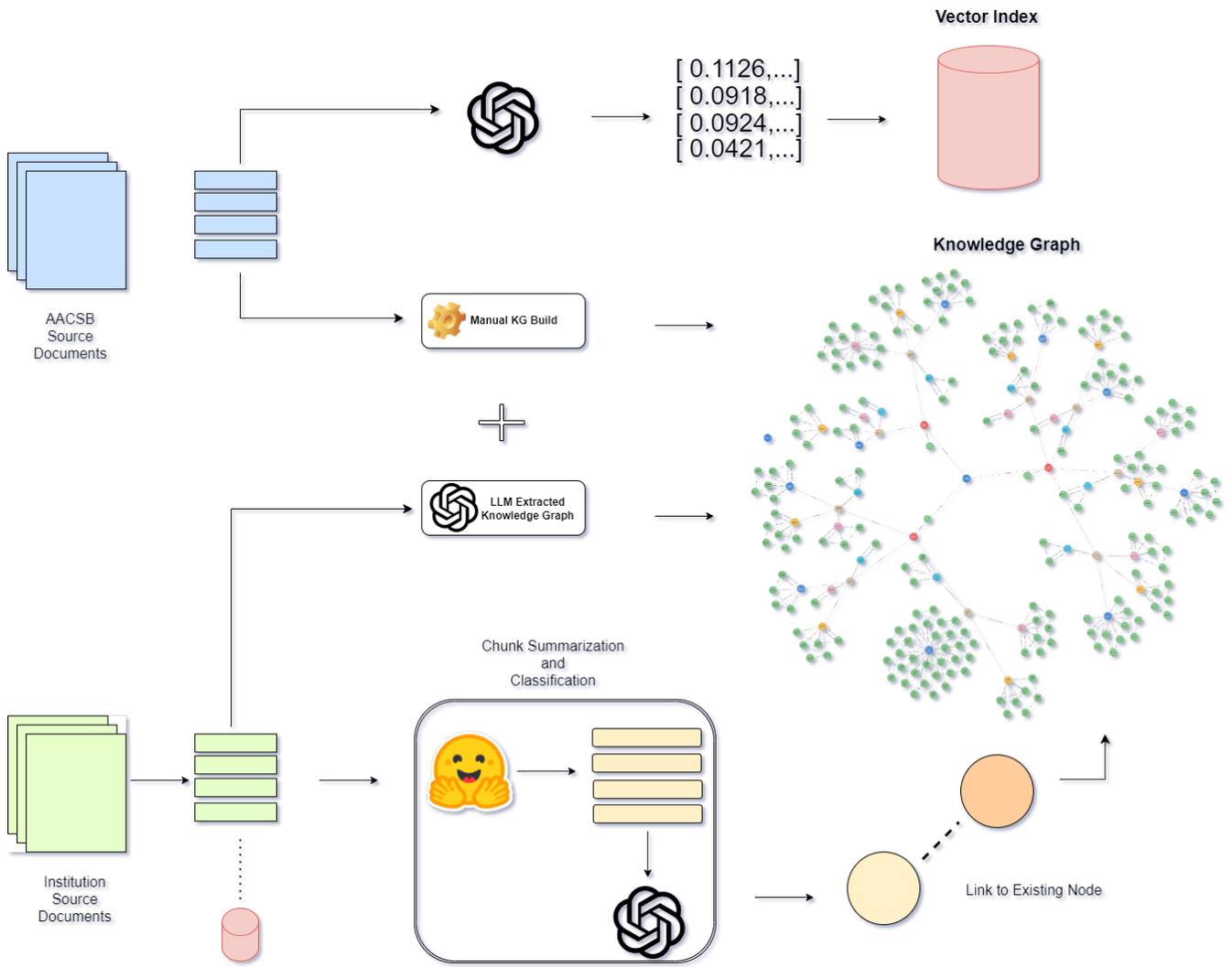

Figure 3: Knowledge Graph Build  Phase 1:

Figure 4: Knowledge Graph Build  Phase 2:

Figure 5: Knowledge Graph Build  Phase 3:

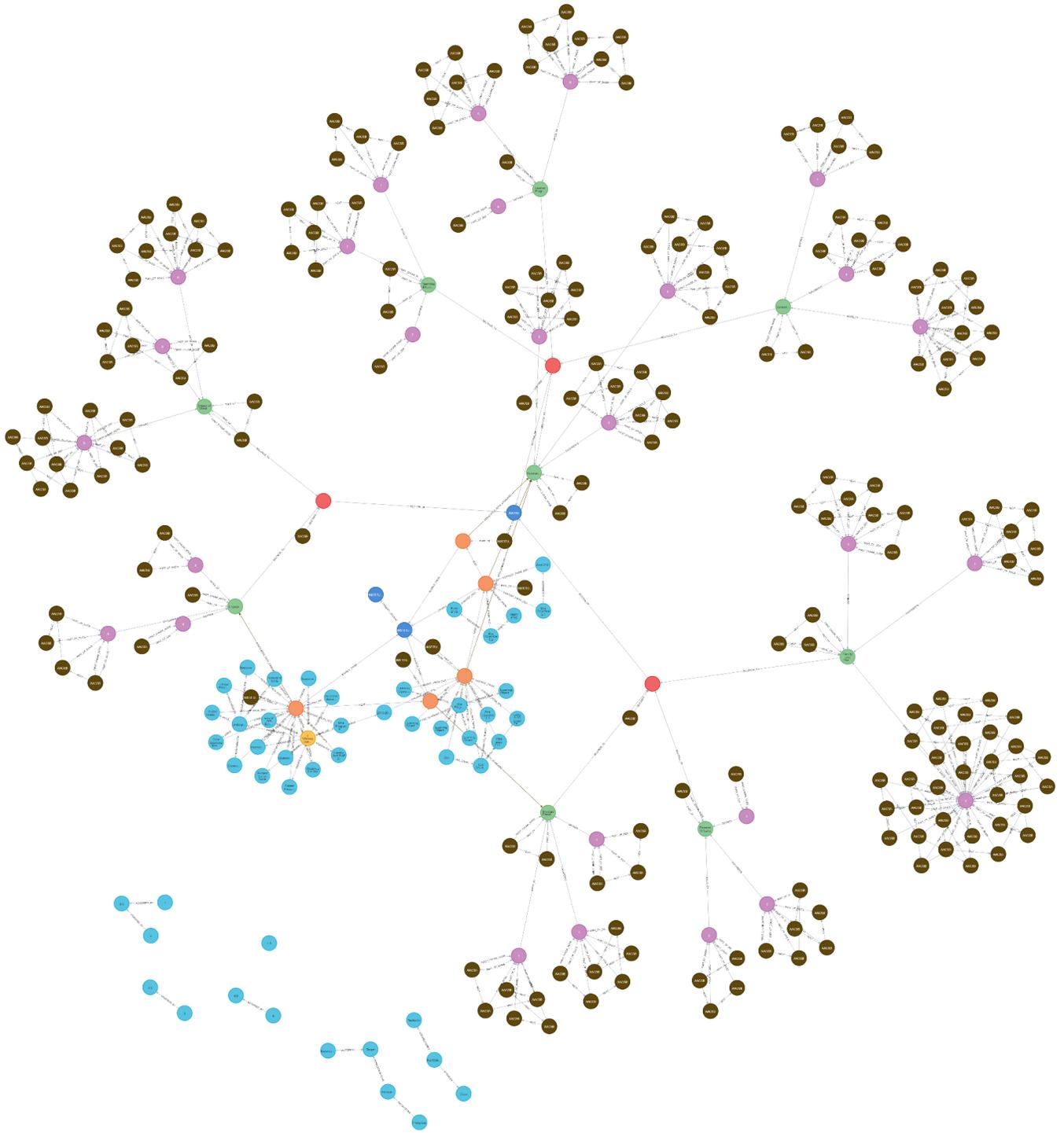